\documentclass[11pt,a4paper]{article}
\usepackage[pdfstartview=FitH,colorlinks=true,linkcolor=blue,anchorcolor=black,citecolor=magenta,urlcolor=blue]{hyperref}
\usepackage[english]{babel}
\usepackage{amsmath,amssymb,titling,authblk}
\usepackage{appendix}
\usepackage{epsfig}
\usepackage{cite}

\allowdisplaybreaks[4]
\numberwithin{equation}{section}

\usepackage {color}

\newcommand{\del}{\partial}
\def\nn{\nonumber}

\newcommand{\be}{\begin{equation}}
\newcommand{\ee}{\end{equation}}
\newcommand{\beqa}{\begin{eqnarray}}
\newcommand{\eeqa}{\end{eqnarray}}
\newcommand{\eqn}[1]{(\ref{#1})}
\newcommand{\R}{\mathbb{R}}
\newcommand{\C}{\mathbb{C}}
\newcommand{\Tr}[1]{\:{\rm Tr}\,#1}

\renewenvironment{thebibliography}[1]
         {\section*{References}\frenchspacing\small
          \begin{list}{[\arabic{enumi}]}
         {\usecounter{enumi}\parsep=2pt\topsep 0pt
         \settowidth{\labelwidth}{[#1]}
         \leftmargin=\labelwidth\advance\leftmargin\labelsep
         \rightmargin=0pt\itemsep=1pt\sloppy}}{\end{list}}

\begin{document}
\setlength{\droptitle}{-6pc}
\title{The Mass Hyperboloid as a Poisson-Lie Group}
\renewcommand\Affilfont{\itshape}
\setlength{\affilsep}{1.3em}

\author[1]{S. G. Rajeev\thanks{s.g.rajeev@rochester.edu}}
\author[2,3]{Patrizia Vitale\thanks{patrizia.vitale@na.infn.it}}
\affil[1]{Department of Physics and Astronomy, Department of Mathematics,University
of Rochester, Rochester, NY 14627, USA}
\affil[2]{Dipartimento di Fisica ``Ettore Pancini'', Universit\`{a} di Napoli {\sl Federico~II}, Napoli, Italy}
\affil[3]{INFN, Sezione di Napoli, Italy}

\date{}

\maketitle
\begin{abstract}
The light cone formalism of a massive scalar field has been shown
by Dirac to have many advantages. But it is not manifestly Lorentz
invariant. We will show that this is a feature not a bug: Lorentz
invariance is indeed a symmetry, but in a different sense defined by Drinfel'd.
The key idea is that the mass shell (mass hyperboloid) is a Poisson-Lie
group: there is a non-abelian group multiplication and non-zero Poisson
brackets between components of four-momentum. Rotations form the dual
group of the hyperboloid in the sense of Drinfel'd. Infinitesimal Lorentz
transformations form a Lie bi-algebra.
\end{abstract}




\section{Introduction}
Quantum groups, originally introduced in relation with quantum integrability more than thirty years ago (see for example \cite{qint}), have acquired a central role in modern literature  as  appropriate symmetries of non-commutative space-time, non-commutative dynamics  and non-commutative models of matter and gauge fields. All this activity is  relevant in view of a consistent theory of quantum gravity which would  imply noncommutative models of spacetime as effective theories, at scales which are large in comparison with   Planck length. 
Poisson-Lie groups are semi-classical approximations of quantum groups,  where non-commutativity of the algebra of functions on  the groups manifold is replaced by non-zero Poisson brackets. As such, they are mainly studied in the same  perspective as above, namely to grasp hints on the deviation from the standard pseudo-Riemannian framework, which is foreseen by most of the QG models on the market. 

In this contribution, we wish to show that these structures do not necessarily need to depart from standard physics in order to be detected, but may play a role in conventional, undeformed dynamical systems. We will discuss a specific example, the Poisson-Lie group $SU(2)$ and its Drinfel'd double \cite{drinfeld, semenov} $SL(2,\C)$, namely the (double covering) Lorentz group,  and show that they are intimately related with the light-front formalism of particle physics, being indeed the most natural framework where to describe the mass hyperboloid of massive scalar fields. The example will give us the opportunity to make explicit the geometric structures of Poisson-Lie  groups, such as Poisson-Lie brackets, dressing actions, Poisson-Lie duality and to place them into a familiar context.

\section{The Hyperboloid}

The four-momentum of a particle with unit mass belongs to one-sheet
of a hyperboloid. There are equivalent descriptions of this manifold:
\[
p_{0}^{2}-p_{1}^{2}-p_{2}^{2}-p_{3}^{2}=1,\quad p_{0}>0\iff
p_{+}p_{-}-(p_{2}^{2}+p_{3}^{2})=1,\quad p_{+}>0
\]
where $p_{+}=p_{0}+p_{1},p_{-}=p_{0}-p_{1}$. We can solve for
\[
p_{-}=\frac{1+p_{2}^{2}+p_{3}^{2}}{p_{+}},\quad p_{0}=\frac{1+p_{+}^{2}+p_{2}^{2}+p_{3}^{2}}{2p_{+}},\quad p_{1}=\frac{-\left(1+p_{2}^{2}+p_{3}^{2}\right)+p_{+}^{2}}{2p_{+}}
\]
in terms of $p_{+},p_{2},p_{3}$. Let us refer to the latter as ${\mathcal P}_+$.  Thus the hyperboloid ${\mathcal P}_+$ can be identified
with the half-space $\mathbb{R}^{+}\times\mathbb{R}^{2}\ni(p_{+},p_{2},p_{3})$.  Notice that, 
if we were to use $p_{1},p_{2},p_{3}$ as independent variables, solving
for $p_{0}$ would involve a square root. This is a disadvantage:
we would have to somehow project out the negative energy solutions.
Dirac was the first to point the convenience of the light--cone (more accurately light--front) framework for  field theory.

Let $\mathcal{H}=\left\{ A\mid A^{\dagger}=A,\ A>0,\ \det A=1\right\} $ be the set of $2\times 2$  positive Hermitian matrices with unit determinant.  There is  a one-to-one  correspondence between ${\mathcal P}_+$ and $\mathcal{H}$. In light-front coordinates the correspondence goes as follows
\be
{\mathcal P}_+\ni (p_+,p_2,p_3)\rightarrow {\mathcal P}(p)=
\left(
\begin{array}{cc}
p_+&p_2+ip_3\\
p_2-ip_3&p_{-}
\end{array}
\right)\in \mathcal{H}
\ee
with $p_+ p_- -(p_2^2+p_3^2)=1$. Moreover, 
a positive Hermitian matrix with unit determinant has a unique decomposition in terms of upper triangular matrices with real diagonal and unit determinant (Cholesky factorisation), ${\mathcal P}(p)=\ell^{\dagger}\ell$ with
\be\label{lambda}
\ell=\frac{1}{\sqrt{p_+}}\left(
\begin{array}{cc}
p_+&p_2+ip_3\\
0&1
\end{array}\right).
\ee
The latter form a group, which we shall refer to as $SB(2,C)$, it being the Borel subgroup of $SL(2,\C)$, as we shall explain in a moment. Therefore, the Cholesky ``square root" of a positive Hermitian matrix with unit determinant is a Lie group. 

Let us remark that:
\begin{itemize}
\item
The Lorentz group acts on ${\mathcal P}(p)\in \mathcal{H}$ by ${\mathcal {\mathcal P}}(p)\mapsto\Lambda {\mathcal P}(p)\Lambda^{\dagger},\Lambda\in SL(2,C)$;
\item
 The subgroup $SU(2)$ acts with $1$ as fixed point, i.e., rotations;
\item The trace of ${\mathcal P}(p)$ is invariant under $SU(2)$:  $\Tr {\mathcal P}(p)= 2 p_0$,  twice the
energy.
\end{itemize}

Besides the standard realization  of the Lie algebra of the Lorentz group, which we shall identify with $\mathfrak{sl}(2,C)$ from now on,  in terms of rotations and boosts, it is possible to choose a different basis, which  emphasizes its bi-algebra structure. Let us indicate with 
$e_a, f^a$ dually related generators with respect to  the
inner product $\mathrm{Im}\Tr$.   There are two  maximally isotropic, non-commuting
sub-algebras, $\mathfrak{su}(2)$ (spanned by $e_{a}$) and the  Borel algebra, $\mathfrak{sb}(2,C)$ (spanned
by $f^{a}$).  The latter is the Lie algebra of the group  $SB(2,C)$. As a vector space $\mathfrak{sl}(2,\C)=  \mathfrak{su}(2)\oplus \mathfrak{sb}(2,\C)$. All together they form a Manin triple.  As a basis we may  choose
\be\label{generators}
\begin{array}{ccc}
e_{1}=\left(\begin{array}{cc}
\frac{i}{2} & 0\\
0 & -\frac{i}{2}
\end{array}\right),& e_{2}=\left(\begin{array}{cc}
0 & \frac{i}{2}\\
\frac{i}{2} & 0
\end{array}\right),&e_{3}=\left(\begin{array}{cc}
0 & -\frac{1}{2}\\
\frac{1}{2} & 0
\end{array}\right)
\\
\\
f^{1}=\left(\begin{array}{cc}
1 & 0\\
0 & -1
\end{array}\right),& f^{2}=\left(\begin{array}{cc}
0 & 2\\
0 & 0
\end{array}\right),& f^{3}=\left(\begin{array}{cc}
0 & 2i\\
0 & 0
\end{array}\right)
\end{array}
\ee
which verify
$
\mathrm{Im\ Tr}\, e_{a}f^{b}=\delta_{b}^{a}
$
with Lie brackets
\be\label{sl1}
[e_{i},e_{j}]={\epsilon_{ij}}^k e_k, \;\; \left[f^{i},f^{j}\right]={f^{ij}}_k f^k,\;\;
 \left[e_i,f^j\right]= {\epsilon_{ki}}^j f^k+ {f_i}^{jk}e_k 
\ee
and ${f_i}^{jk}= 2\epsilon_{1i\ell}\epsilon^{\ell jk}$ the structure constants of the Lie algebra $\mathfrak{sb}(2,\C)$. Notice the mutual adjoint action of one subalgebra onto the other in the last Lie bracket.

 The corresponding dual subgroups of $SL(2,\C)$, which realize the latter as a Drinfel'd double, are $SU(2)=\left\{ u\mid u^{\dagger}u=1,\ \det u=1\right\} $
and $SB(2,\C)=\left\{ \lambda=\left(\begin{array}{cc}
a & b+ic\\
0 & a^{-1}
\end{array}\right),a>0,b+ic\in\mathbb{C}\right\} $. Therefore an element of the Lorentz group $\gamma\in SL(2,\C)$, may be locally parametrized as a product $\gamma=u\cdot \lambda$ or alternatively  $\gamma=\lambda'\cdot u'$. Notice that this is nothing but   the familiar Iwasawa decomposition of $SL(2,\C)$, $g= k \cdot a\cdot  n$, with 
$k\in SU(2), \,a\cdot n= b \in SB(2,\C)$ (see \cite{sawyer} for a review). 


\section{The Hyperboloid  Group}\label{semidir}
As we have seen, the hyperboloid ${\mathcal P}_+$ is itself a group, it being $\ell\in SB(2,\C)$. Therefore $SB(2,\C)$ plays  multiple roles:  on one hand it  is a factor of the Lorentz group in the Iwasawa decomposition,   and, as such, it acts  with the appropriate representation;   on the other hand, it models  the manifold of momenta  for particles of unit mass, therefore, it is acted upon by the Lorentz group and its factors; moreover, it being the dual group of $SU(2)$, its group elements replace standard angular momentum which is usually associated with the Abelian dual Lie algebra of $\mathfrak{su}(2)$ through momentum map. Let us see these structures in more detail.

 The left-invariant Maurer-Cartan form can be easily calculated: 
 \be
 \alpha= \lambda^{-1} d\lambda = \frac{da}{a} f^1 + \frac{1}{2} \left(\frac{db}{a}+ \frac{b}{a^2} da\right) f^2 +  \frac{1}{2} \left(\frac{dc}{a}+ \frac{c}{a^2} da\right) f^3
 \ee 
 This yields, under the identification of $\lambda\rightarrow \ell$, namely $a\rightarrow \sqrt{p_+}, \, b\rightarrow \frac{p_2}{\sqrt{p_+}},\, c\rightarrow \frac{p_3}{\sqrt{p_+}}$, a basis of  left-invariant one-forms
 \be\label{tilsbforms}
 \alpha_1= \frac{1}{2}\frac{d p_+}{p_+}, \;\; \alpha_2 =\frac{1}{2} \frac{dp_2}{p_+} , \;\; \alpha_3=   \frac{1}{2} \frac{dp_3}{p_+} 
 \ee
%
 with left-invariant vector fields
 \be\label{tilsbvfields}
{Y}^1= 2 p_+\frac{\del}{\del p_+},\;\;\; {Y^2}= 2p_+ \frac{\del}{\del p_2},\;\; \;{Y^3}= 2 p_+\frac{\del}{\del p_3}
\ee
satisfying $ [Y^i, Y^j]= {f^{ij}}_k Y^k$.
The latter generate the right action of $SB(2,\C)$ on the momentum hyperboloid. The left action is obtained by repeating the above for the right-invariant Maurer-Cartan form. We get the right-invariant forms
\be\label{tilsbformsright}
\theta_1= \frac{1}{2}\frac{d p_+}{p_+}, \;\; \theta_2 =\frac{1}{2} \left(dp_2-\frac{p_2}{p_+} dp_+\right), \;\; \theta_3=   \frac{1}{2} \left(dp_3-\frac{p_3}{p_+} dp_+\right)
 \ee
 with right-invariant vector fields
 \be\label{tilsbvfieldsright}
{Z}^1= 2\left( p_+\frac{\del}{\del p_+}+p_2\frac{\del}{\del p_2}+p_3\frac{\del}{\del p_3}\right)
,\;\;\; {Z^2}= 2 \frac{\del}{\del p_2},\;\; \;{Z^3}= 2 \frac{\del}{\del p_3}
\ee
satisfying $[{Z}^i, {Z}^j]= - f^{ij}_k  Z^k$.
\subsection{Poisson-Lie  brackets}\label{Poiliebr}
In terms of  a generalisation of the Kirillov-Souriau-Konstant bracket (KSK), which is naturally defined on the dual of any Lie algebra,  dually related Lie groups possess a natural Poisson bracket (that reduces to the KSK bracket when evaluated at the identity), which is compatible with the group action, namely the group multiplication results to be  a Poisson morphism. It has been shown (see \cite{chari} for a review) that  Poisson-Lie brackets for a given Lie group $G$ may always been written as 
\be\label{Liepoi}
\{\gamma_1,\gamma_2\}= [r, \gamma_1\gamma_2].
\ee
with $\gamma\in G$ and  $r\in \mathfrak{g}\otimes\mathfrak{g}$  the so called
  classical $r$-matrix, solution of  the classical (possibly modified) Yang-Baxter equation (CYB) \cite{chari}. It is standard notation to denote $\gamma_1= \gamma\otimes 1, \gamma_2= 1\otimes \gamma$ with $\gamma_1\gamma_2= \gamma\otimes\gamma$. The Lorentz group itself is a Poisson-Lie group with  
  \be r= e_i\otimes f^i, \;\;\;\;\; [r_{12}, r_{13}+ r_{23}]+[r_{13}, r_{23}]=0 \;\;\rm{(CYB)}.
  \ee
 When specialised to $\lambda\in SB(2,\C)$ Eq. \eqn{Liepoi} yields
 \beqa
 \{\lambda\stackrel{\otimes}{,}\lambda\}&=& e_i \lambda\otimes f^i \lambda-\lambda e_i\otimes\lambda  f^i \\
 \{\lambda\stackrel{\otimes}{,}\lambda^\dagger\}&=&  e_i \lambda \otimes (f^i\lambda)^\dagger -\lambda e_i\otimes(\lambda  f^i)^\dagger
 \eeqa
 namely
 \be\label{casim}
 \{a,b\}= ac, \;\;\;  \{a,c\}= -ab,\;\;\;  \{b,c\}= a^2-\frac{1}{a^2}
 \ee
 The latter may be rewritten in terms of the hyperboloid coordinates to give
 \be\label{casim}
 \{p_+, p_2\}= 2p_+ p_3, \;\;\;   \{p_3, p_+\}= 2p_+ p_2,\;\;\;  \{p_2,p_3\}= p_+^2-p_2^2-p_3^2-1
 \ee
 The Poisson algebra \eqn{casim} admits a Casimir function
 which corresponds to the energy $p_0$,  
 $
 C= \frac{1}{2}\left(p_++\frac{1}{p_+}+\frac{p_2^2+p_3^2}{p_+}\right).
$
Notice however that the group $SB(2,C)$ has no Casimir intended as a polynomial in the center of its universal enveloping algebra (see for example \cite{selene}, where $\mathfrak{sb}(2,\C)$  is classified as a type B algebra).

\subsection{Dressing action of $SU(2)$ on the hyperboloid}\label{su2dress}
The Drinfel'd double $SL(2,C)$ is  a deformation of the semi-direct product $SU(2)\ltimes \R^3$, with $\R^3$ the dual algebra of $SU(2)$. In standard Poisson geometry the momentum map associated with the Hamiltonian action  of $SU(2)$ on $\R^3$ gives rise to Hamiltonian functions which close the algebra of $SU(2)$ with respect to the KSK Poisson bracket. In classical mechanics  these  functions are the three components of the angular momentum, $\ell_i$,  according to $\Lambda(d\ell_i) =- V_i$, where $\Lambda$ is the KSK Poisson tensor and $V_i$ the vector fields realizing the infinitesimal action of $SU(2)$ on $\R^3$. 

The generalization of the above construction to the Drinfel'd double $SL(2,\C)$ yields the infinitesimal dressing action of $SU(2)$ on $SB(2,\C)$. 
The Maurer-Cartan forms $\alpha^i$ derived in Eq. \eqn{tilsbforms} play now the role of the exact one-forms $d\ell_i$ whereas $\Lambda$ is replaced by  the Poisson-Lie tensor computed in Sec. \ref{Poiliebr}, 
\be\label{PoiLieP}
P= 2p_+p_3 \frac{\del}{\del p_+}\wedge \frac{\del}{\del p_2} -2p_+p_2 \frac{\del}{\del p_+}\wedge \frac{\del}{\del p_3}- (1-p_+^2+p_2^2+ p_3^2) \frac{\del}{\del p_2}\wedge \frac{\del}{\del p_3}. 
\ee
This yields
$
\;L_{i}=-P(\alpha_i) ,\;
$
that is, in the light-front coordinates of the hyperboloid
\be\label{rightdressu2}
\begin{array}{lll}
L_1&=& -p_2\frac{\del}{\del p_3} + p_3 \frac{\del}{\del p_2}\\
L_2&=& -p_3\frac{\del}{\del p_+} -\frac{1}{2 p_+}(1-p_+^2+p_2^2+p_3^2)  \frac{\del}{\del p_3}\\
L_3&=& p_2\frac{\del}{\del p_+} +\frac{1}{2 p_+}(1-p_+^2+p_2^2+p_3^2)  \frac{\del}{\del p_2}
\end{array}
\ee
They satisfy the commutation relations of the Lie algebra $\mathfrak{su}(2)$, $[L_i, L_j]= {\varepsilon_{ij}}^k L_k$. That is, they generate the dressing action
of $SU(2)$ on the hyperboloid.
{The latter  is the dressing action generated through left-invariant one-forms. Analogously, we have another one, generated by right-invariant forms \eqn{tilsbformsright}. We have 
$ \;
\widehat{L}_{\ell}=-P(\theta_\ell) ,\;
$
that is, in the light-front coordinates of the hyperboloid
\be\label{leftdressu2}
\begin{array}{lll}
\widehat{L}_1&=& p_3\frac{\del}{\del p_2} - p_2 \frac{\del}{\del p_3}\\
\widehat{L}_2&=&- p_3(p_+\frac{\del}{\del p_+}+p_2\frac{\del}{\del p_2}) -\frac{1}{2}(1-p_+^2-p_2^2+p_3^2)  \frac{\del}{\del p_3}\\
\widehat{L}_3&=& p_2(p_+\frac{\del}{\del p_+} +p_3\frac{\del}{\del p_3})+\frac{1}{2}(1-p_+^2+p_2^2-p_3^2)  \frac{\del}{\del p_2}
\end{array}
\ee 
They close the same $SU(2)$ algebra, $ [\widehat{L}_i, \widehat{L}_j]=  {\varepsilon_{ij}}^k \widehat{L}_k$.} 

\section{ The group manifold of $SU(2)$ }
As well as the group manifold of $SB(2,\C)$, which we have regarded as the mass hyperboloid, the group manifold $SU(2)$, namely the three-sphere $S^3$, is an interesting configuration space for classical dynamics(see for example the rigid rotor and all its applications, or the non-linear sigma model with target space the sphere $S^3$). 
Therefore it is worth to duplicate  the previous construction, where now the role of the two groups is exchanged. 

A convenient parametrization of $u\in SU(2)$ is $u= y^0 \mathbf{1} +  2 y^i e_i$ with $y^\mu\in \R^4, \sum y^\mu y^\mu = 1$  and $e_i$ the generators of $SU(2)$.  given by the first line of  Eq. \eqn{generators}. 
Explicitly
\be
u=\left(
\begin{array}{cc}
y^0 + i y^1& i y^2 - y^3\\
i y^2 + y^3,&y^0 - i y^1
\end{array}\right)\;\;\;\; {\rm with}\; \det u=1
\ee
From the  left-invariant Maurer-Cartan one-form $u^{-1}du$  the three basis left-invariant one-forms are obtained
\be\label{etaleft}
\eta^i= 2\left(y^0 dy^i-y^i dy^0-{\varepsilon^i}_{jk}y^j dy^k\right)
\ee
with dual vector fields 
\be\label{Wleft}
W_i= \frac{1}{2}\left(y^0\del_i-y^i\del_0-{\varepsilon_{ij}}^k y^j\del_k\right)
\ee
Analogously, from $du \,u^{-1}$ the right-invariant forms  are found to be
\be\label{omegaright}
\omega^i= 2\left(y^0 dy^i-y^i dy^0+{\varepsilon^i}_{jk}y^j dy^k\right).
\ee
with dual vector fields 
\be\label{Vright}
V_i= \frac{1}{2}\left(y^0\del_i-y^i\del_0+{\varepsilon_{ij}}^k y^j\del_k\right)
\ee

\subsection{Poisson-Lie bracket on $SU(2)$}
The Poisson-Lie  bracket \eqn{Liepoi} is defined on the whole group $SL(2,\C)$. For
\be
SL(2,\C)) \ni \gamma=\left(
\begin{array}{cc}
\alpha&\beta\\
\delta&\zeta
\end{array}\right)
\ee
with  $\alpha\zeta-\beta\delta=1$, $\alpha,\beta,\delta,\zeta$ complex variables we obtain 
\beqa\label{sl2bra}
\{\alpha,\beta\}=-i \alpha\beta,\;\;&\{\alpha,\delta\}=-i \alpha\delta,\;\;&\{\alpha,\zeta\}=-2i \beta\delta\nn\\
\{\beta,\delta\}=0,\;\;\;\;\;\;\;\;\;&\{\beta,\zeta\}=-i\beta\zeta,\;\;&\{\delta,\zeta\}=-i\delta\zeta\,.
\eeqa
Specialising it to  $u\in SU(2)$, we find 
\beqa
\{y_0,y_1\}=-(y_2^2+y_3^2),\;\;&\{y_0,y_2\}=y_1 y_2,\;\;\;\;&\{y_0,y_3\}=y_1 y_3\nn
\\
\{y_1,y_2\}=-y_0 y_2,\;\;\;\;\;\;\;\;\;&\{y_1,y_3\}=-y_0 y_3,\;\;&\{y_2,y_3\}=0\,\label{Poisu2}
\eeqa
The Poisson  tensor which endows $SU(2)$ with a Poisson-Lie structure is therefore
\be\label{PoiLieQ}
Q= (y_1\del_0-y_0\del_1)\wedge(y_2\del_2+y_3\del_3) -(y_2^2+y_3^2) \del_0\wedge\del_1.
\ee
\subsection{Dressing actions of $SB(2,\C)$ on $SU(2)$}
By repeating the calculation performed in Sec. \ref{su2dress} for the group $SB(2,\C)$ it is possible to get the dressing actions (left and right) of the latter on $SU(2)$.
In comparison with the standard momentum map picture, the left-invariant Maurer-Cartan one-forms $\eta^i$  (resp. right-invariant) derived above play now the role of the exact one-forms $d\ell_i$ whereas the Konstant-Souriau-Kirillov bracket $\Lambda$ is replaced by  the Poisson-Lie tensor \eqn{PoiLieQ}. The generators of the right dressing action (respectively left dressing action) of $SB(2,\C)$ on $SU(2)$ are therefore retrieved according to 
$\; B^{\ell}=-Q(\eta^\ell), \;
$
which gives 
\be
B^1=-2(y_2\del_2+y_3\del_3),\;\; B^2= 2(y_2\del_1-y_3\del_0),\;\; B^3= 2(y_2\del_0+y_3\del_1)
\ee
where use has been made of the constraints on the parameters of $SU(2)$, $y_\mu y^\mu=1$ implying  $y^\mu\del_\mu= 0$. 
They satisfy the commutation relations of the Lie algebra $\mathfrak{sb}(2)$, 
\be
[B^1,B^2]=-2 B^2,\;\; [B^1,B^3]=-2 B^3, \;\; [B^2,B^3]=0
\ee
{Analogously, from the   right-invariant one-forms
\be
\omega^i= 2\left(y^0 dy^i-y^i dy^0+{\varepsilon^i}_{jk}y^j dy^k\right).
\ee
We get the vector fields $\;
\widehat{B}^{\ell}=-Q(\omega^\ell),\;$ so that 
\be\label{hatBri}
\widehat{B}^1= B^1,\;\; \widehat{B}^2=2(y_2\del_1+y_3\del_0),\;\;\widehat{B}^3=2(y_3\del_1-y_2\del_0)
\ee
which close the same Lie algebra as the previous one, therefore, they generate the left dressing action of $SB(2,\C)$ on $SU(2)$.} 
\section{Dressing actions of  $SL(2,\C)$ on the hyperboloid}
In order to obtain the action of the Lorentz group on the mass hyperboloid, it is natural to exploit the isomorphism of its Lie algebra  with $\mathfrak{sl}(2,\C)$ and look for its dressing action on the dual.
The group $SL(2,\C)$ is itself a Poisson-Lie group with dual group another copy of $SL(2,\C)$, which shall be indicated by $SL(2,\C)^*$ to keep track of the doubling (and to stress that it is the exponentiation of the dual Lie algebra). Therefore, the Poisson-Lie bracket   \eqn{Liepoi} may be used to derive {the left and right dressing actions of $SL(2,\C)$ on its dual. }
The hyperboloid (identified with the group manifold of $SB(2,\C)$) may be regarded as the quotient $SL(2,\C)^*/SU(2)$. {Depending on the decomposition chosen, namely $SL(2,\C) \sim SB(2,\C) \cdot SU(2)$ or $SL(2,\C) \sim SU(2) \cdot SB(2,\C)$, the hyperboloid is  a right coset or a left coset.\footnote{Dually, $SU(2)$ is a left or right quotient with respect to $SB(2,\C)$. This remark and the next are relevant when, as in previous section, it is the group manifold of $SU(2)$ which is associated with the carrier space of dynamics, and, therefore, one is interested in the action of the Lorentz group on it.  } 
Therefore, once a decomposition chosen, the infinitesimal generators of the dressing action of $SL(2,\C)$  on its dual have to be computed consistently. In this way, they  will pass to the quotient and give  
 the appropriate (left or right) dressing action of the whole Lorentz group on the hyperboloid.\footnote{ Analogously, the appropriate dressing action of the Lorentz group on $SU(2)$ is obtained. }}

First of all, we choose to represent   $\gamma\in SL(2,\C)$ as the product $\gamma = \lambda u$, with $\lambda\in SB(2,\C)$ and $u\in SU(2)$,  parametrized as above. Then, in order to write  the Poisson-Lie bi-vector field  in the chosen parametrization,  we compute the LHS of \eqn{Liepoi}
$$
\begin{array}{ll}
&\{\gamma_1,\gamma_2\}= \lambda_1\{u_1, \lambda_2\} u_2+ \lambda_1\lambda_2\{u_1, u_2\}+\{\lambda_1,\lambda_2\}u_1 u_2+\lambda_2\{\lambda_1,u_2\} u_1\\
&= \lambda_1\{u_1, \lambda_2\} u_2+ \lambda_1\lambda_2 (r u_1 u_2-u_1 u_2 r) +
 (r \lambda_1\lambda_2-\lambda_1\lambda_2 r) u_1 u_2+\lambda_2\{\lambda_1,u_2\}u_1
\end{array}
$$
where use has been made of the Poisson brackets for the $SU(2)$ and $SB(2,\C)$ variables, $ \{u_1, u_2\} = [r, u_1u_2]$ and $ \{\lambda_1, \lambda_2\} = [r, \lambda_1\lambda_2]$. The RHS of \eqn{Liepoi} yields instead
\be
r \gamma_1\gamma_2-\gamma_1\gamma_2 r= r \lambda_1 u_1\lambda_2 u_2-\lambda_1 u_1\lambda_2 u_2 r\,.
\ee
 Comparing the two we get
$\;
\lambda_1\{u_1, \lambda_2\} u_2 +\lambda_2\{\lambda_1,u_2\}u_1= 0, \;
$
from which, by consistency,
$$
\{u_1,\lambda_2\}=\lambda_2 r_{12} u_1= e_i u \otimes \lambda f^i ,\;\;\; 
 \{\lambda_1, u_2\}= -\lambda_1r_{12} u_2=-\lambda e_i\otimes f^i u.
$$
Explicitly they yield  \small{
\be\label{mixedpoi}
\begin{array}{llll}
\{y_0, a\}=-\frac{a}{2}y_1 & \{y_1, a\}=\frac{a}{2}y_0& \{y_2, a\}=-\frac{a}{2}y_3 & \{y_3, a\}= \frac{a}{2}y_2 \\
\{y_0, b\}=\frac{b}{2}y_1-a y_2 &  \{y_1, b\}=-\frac{b}{2}y_0+a y_3&  \{y_2, b\}=\frac{b}{2}y_3+a y_0 & \{y_3, b\}=-\frac{b}{2}y_2-a y_1\\
 \{y_0, c\}=\frac{c}{2}y_1-a y_3 &  \{y_1, c\}= -\frac{c}{2}y_0-a y_2 
& \{y_2, c\}=\frac{c}{2}y_3+a y_1 &\{y_3, c\}= -\frac{c}{2}y_2+a y_0 
\end{array}
\ee}
 Therefore the Poisson-Lie tensor for $SL(2,\C)$ may be written as $\;
 \mathbb{P}= P+ Q+ M, \;$
 with $P$  and $Q$  respectively given by Eq. \eqn{PoiLieP} and   Eq. \eqn{PoiLieQ}, while $M$ is to be read from \eqn{mixedpoi}, that is
\be\label{PoiLieM}
\begin{array}{lllll}
M&= &\left[\frac{a}{2} y_1 \del_a+(ay_2-\frac{b}{2}y_1)\del_b + (a y_3-\frac{c}{2}y_1)\del_c\right]&\wedge& \del_{y_0} \\
&- &\left[\frac{a}{2} y_0 \del_a+(ay_3-\frac{b}{2}y_0)\del_b - (a y_2+\frac{c}{2}y_0)\del_c\right]&\wedge& \del_{y_1} \\
&+&\left[\frac{a}{2} y_3 \del_a-(ay_0+\frac{b}{2}y_3)\del_b - (a y_1+\frac{c}{2}y_3)\del_c\right]&\wedge& \del_{y_2} \\
&-&\left[\frac{a}{2} y_2 \del_a-(ay_1+\frac{b}{2}y_2)\del_b +(a y_0-\frac{c}{2}y_2)\del_c\right]&\wedge& \del_{y_3} 
\end{array}
\ee
Notice that, for simplicity, Eqs. \eqn{mixedpoi} and \eqn{PoiLieM} are not explicitly written in the light-front coordinates of the hyperboloid. We shall see, however, that the  parametrization   will homogenize after projection. 

We now observe that a basis of left- (resp. right-) invariant one-forms for $SL(2,\C)$ is represented by $(\eta^i, \alpha_i)$ (resp. $(\omega^i, \theta_i)$). 
Therefore we may compute the dressing action of the  Lorentz group on its dual by following the same procedure (the generalised momentum map) as in previous sections. Then, by considering the quotient with respect to each sub-group we shall obtain the dressing action of the Lorentz group on each of its factors.

Notice that, having chosen the parametrization $\gamma = \lambda u$, the group $SB(2,\C)$ is a right coset with respect to the $SU(2)$ action, while $SU(2)$ is a left coset with respect to the $SB(2,\C)$ action. Therefore, we have to consider the pair  $(\omega^i, \alpha_i)$ namely, right-invariant forms of $SU(2)$ and left-invariant forms of $SB(2,\C)$ in order to retrieve the dressing action correctly, upon projection. Let us start with $\mathbb{P}(\omega^i)$. We have to evaluate
\be\label{Pomega}
\mathbb{P}(\omega^i) =P(\omega^i) + Q(\omega^i) + M(\omega^i)
\ee
$P(\omega^i)= 0$ trivially while $Q(\omega^i)= - \hat B^i$ according to Eq. \eqn{hatBri}, Eqs. \eqn{hatBri1}-\eqn{hatBri3}, namely the generators of the left dressing action of $SB(2,\C)$ on $SU(2)$.
Moreover,  
$\; 
M(\omega^i)= - Y^i, \;$
with $Y^i$ the left-invariant vector fields of $SB(2,\C)$ already computed in \eqn{tilsbvfields}. Namely,
$
\mathbb{P}(\omega^i) = - \hat B^i - Y^i, \;
$
yielding in turn 
\be\label{promega}
\pi_*^R\mathbb{P}(\omega^i) = - Y_i,\;\;\; \pi_*^L\mathbb{P}(\omega^i) = - \hat B_i
\ee
where we have indicated with  $\pi^R, \pi^L$respectively the projection of $SL(2,\C)$ to the  quotients $SL(2,\C)/SU(2)$, $SL(2,\C)/SB(2,\C)$.
Analogously,
\be
\mathbb{P}(\alpha_i)= P(\alpha_i) + Q(\alpha_i) + M(\alpha_i)
\ee
with now $Q(\alpha_i)=0$ trivially, because $Q$ is the Poisson-Lie tensor on $SU(2)$ while $\alpha_i\in \Omega^1(SB(2,\C))$. As for $M(\alpha_i)$, we find 
$\;
M(\alpha_i) = y^i \del_0-y^0 \del_i -\varepsilon^i_{jk}y^j \del_k,\;$
that is, $M(\alpha_i)= -V_i$,  the right-invariant vector fields of $SU(2)$ computed in Eq. \eqn{Vright}.
  $P(\alpha_i)$ in turn yields   the generators of the right dressing action of $SU(2)$ on $SB(2,\C)$ that we have already computed with Eq. \eqn{rightdressu2},
$\;
P(\alpha_i)=  L_i.\;$
Namely, 
$\;
\mathbb{P}(\alpha_i) = L_i - V_i, \;  
$
yielding in turn 
\be\label{pralpha}
\pi_*^R\mathbb{P}(\alpha_i) = L_i,\;\;\; \pi_*^L\mathbb{P}(\alpha_i) = - V_i.
\ee
Summarising, by contracting the Poisson-Lie tensor $\mathbb{P}$ of $SL(2,\C)$ with an appropriate  basis of one-forms of its dual, $(\alpha_i, \omega^i) \in \Omega^1(SL(2,\C)^*)$, we have obtained a set of vector fields, with right and left projections given respectively by   $(Y^i, L_i)\in \mathfrak{X}(SB(2,\C))$ and $( \hat B^i, V_i)\in \mathfrak{X}(SU(2))$.

It may be verified that $Y^i, L_i$ verify the Lie bi-algebra relations
\be
[{L}_i, Y^j]=Y^k {\epsilon_{ki}}^j+ {2 \epsilon_{1i\ell} {\epsilon}^{\ell jk} }{L}_k
\ee
besides $\;\;[{L}_i, {L}_j]= {\epsilon_{ij}}^k  L_k, \;\;\; [Y^i,Y^j]= {f^{ij}}_k Y^k, \;\;$
where {${f^{ij}}_k= 2 {\epsilon^{1i}}_\ell{\epsilon^{\ell j}}_k$,} therefore representing the infinitesimal action of the Lorentz group on the hyperboloid. 

Dually, $\hat B_i, V_i$ obey the same bi-algebra relations, but are vector fields on the group manifold of $SU(2)$; therefore, they represent  the infinitesimal action of the Lorentz group on the three sphere.

%

\section{Conclusions}
$SL(2,\C)$, the universal covering of the Lorentz group, is a fantastic arena where to probe our understanding of Poisson-Lie geometry, Manin triples and dually related groups. We have proposed in this short contribution,  with no pretention of rigor, a  hopefully pedagogical approach.

Besides describing in detail the geometric structures involved,  we have singled out two physically interesting applications of the formalism: one is already well known and widely studied, the other is new, to our knowledge. 
The former deals with the identification of the group manifold of $SU(2)$ with the target  space of some dynamical system (such as the rigid rotor,  or  the sigma model with and without Wess-Zumino term, see for example \cite{Rajeev, Marotta:2018swj, Marotta:2019wfq, Bascone:2020dcn}). The latter explores the lesser known group manifold of $SB(2,\C)$, which may be naturally associated with the mass hyperboloid of relativistic particles, in the light-front formalism. This second example might have interesting applications which are still under investigation.

\section*{Acknowledgments}
 This contribution is dedicated to A. P. Balachandran,  on the occasion of his eighty-fifth birthday.



\end{document}